\documentclass[12pt]{article}

\usepackage{sbc-template}
\usepackage{graphicx,url}
\usepackage{float}
\usepackage[utf8]{inputenc}  
\usepackage{macros}
\usepackage{amsmath,amsfonts}
\usepackage{multirow}
\usepackage{multicol}
\usepackage[numbers]{natbib}
\title{\dod: Distributed On-Demand ETL for Near Real-Time Business Intelligence}

\author{Gustavo V. Machado\inst{1}, \'{I}talo Cunha\inst{1}, Adriano C. M. Pereira\inst{1}, Leonardo B. Oliveira\inst{1,2}}

\address{Department of Computer Science, Universidade Federal de Minas Gerais \\
Belo Horizonte -- MG -- Brazil
\nextinstitute
  Visiting Associate rofessor -- Department of Computer Science -- Stanford University\\
  Palo Alto -- California -- U.S.A.
  \email{gustavovm@ufmg.br, cunha@dcc.ufmg.br, adrianoc@dcc.ufmg.br, leob@dcc.ufmg.br}
}

\begin{document} 

\maketitle

\begin{abstract}
The competitive dynamics of the globalized market demand information on the internal and external reality of corporations. Information is a precious asset and is responsible for establishing key advantages to enable companies to maintain their leadership. However, reliable, rich information is no longer the only goal. The time frame to extract information from data determines its usefulness. This work proposes \dod{}, a tool that addresses, in an innovative manner, the main bottleneck in Business Intelligence solutions, the Extract Transform Load process (ETL), providing it in \nrt{}. \dod{} achieves this by combining an on-demand data stream pipeline with a distributed, parallel and technology-independent architecture with in-memory caching and efficient data partitioning. We compared \dod{} with other Stream Processing frameworks used to perform \nrt{} ETL and found \dod{} executes workloads up to 10 times faster. We have deployed it in a large steelworks as a replacement for its previous ETL solution, enabling near real-time reports previously unavailable.
\end{abstract}
     
%

\section{Introduction}
Today, there is a dire need for organizations to find new ways to succeed in an increasingly competitive market. There is no simple answer on how to achieve this goal. One thing is patently true, though: organizations must make use of \nrt{} and reliable information to thrive in the global market. 

Business Intelligence (BI) is a term used to define a variety of analytical tools that provide easy access to information that support decision-making processes \cite{malhotra2001information}. These tools perform collection, consolidation, and analysis of information, enabling analytical capabilities at every level inside and outside a company. Putting it another way, BI allows collected data to be unified, structured, and thus presented in an intuitive and concise manner, assisting organizations in corporate decision-making.

The Extract Transform Load (ETL) pipeline is a vital procedure in the Business Intelligence (BI) workflow. It is the process of structuring data for querying or analysis. ETL is made up of three stages, namely: data extraction, data transformation, and data loading where, respectively, data is extracted from their sources, structured accordingly, and finally loaded into the target data warehouse. Two processing strategies can be used in ETL process: (1) Batch and (2) Stream processing. The difference between them resides in whether the source data is bounded, by known and finite size, or unbounded (arriving gradually over time).


The integration of production systems and BI tools, which is a responsibility of ETL processes, ``is the most challenging aspect of BI, requiring about 80 percent of the time and effort and generating more than 50 percent of the unexpected project costs'' \cite{watson2007current}. For all that, ETL is deemed a mission-critical process in BI and deserves close attention. Getting current, accurate data promptly is essential to the success of BI applications. However, due to the massive amount of data and complex operations, current ETL solutions usually have long run times and therefore are an obstacle to fulfilling BI's requirements.


The main challenges of ETL lie on performance degradation at data sources during data extraction, and on performing complex operations on large data volumes in short time frames. The ideal solution has two conflicting goals: (1) cause no impact on data sources and (2) process data in \nrt{} as they are generated or updated. Ideal solutions should handle high-volume input data rates and perform complex operations in short time frames while extracting data with no operational overhead. Both batch and \nrt{} ETL process and its main characteristics are shown in \strf{batchvsstreametll}.

\begin{figure}[H]
\centering
\includegraphics[scale=0.5]{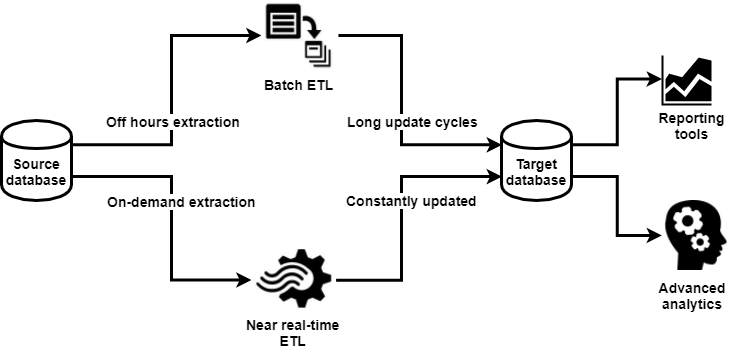}
\caption{Batch vs. Near real-time ETL.}
\label{batchvsstreametll}
\end{figure}

Sabtu et al. \cite{sabtu2017challenges} enumerate several problems related to \nrt{} ETL and, along with Ellis \cite{ellis2014real}, they provide some directions and possible solutions to each problem. However, due to these problems complexity, ETL solutions do not always address them directly: to avoid affecting efficiency on transaction databases, ETL processes were usually run in batches and off-hours (i.e., after midnight or weekends) and, by avoiding peak hours, the impact on mission-critical applications is mitigated. However, in a context where the delay to extract information from data determines its usefulness, BI tools and decision making are heavily impacted when the ETL process is executed infrequently.



In this paper, we propose \dod{}, Distributed On-Demand ETL, a technology-independent tool which combines modern frameworks with custom optimizations and multiple strategies to provide a \nrt{} ETL. \dod{} has minimum impact on the source database during data extraction, delivers a stream of transformed data to the target database at the same speed as data is generated or updated at the source, and provides scalability, being able to respond to data and complexity growth. We achieve all this by synergistically combining multiple strategies and technologies that were once used separately (e.g., in \cite{mesiti2016streamloader,naeem2008event,zhang2015task,jain2012refreshing}): log-based Change Data Capture (CDC), stream processing, cluster computing, an in-memory data store, a buffer to guarantee join consistency along with efficient data partitioning and an unified programming model. \dod{} works in a distributed fashion and on top of a Stream Processing framework, optimizing its performance.



We have developed a \dod{} prototype based on Kafka \cite{kreps2011kafka}, Beam \cite{apachebeam} and Spark Streaming \cite{zaharia2012discretized}. We evaluate \dod's performance executing the same workload on a Stream Processing framework with and without \dod{}. We have found that our solution, indeed, provides better performance when compared to an unmodified stream processing framework, being able to execute workloads up to 10 times faster without losing its core features: horizontal scalability and fault-tolerance. We have also tested it in a large steelworks as a replacement for its previous ETL solution. \dod{} has been able to speed up the ETL process from hours to less than a minute. This, in turn, enabled important \nrt{} reports that were previously impractical.


Our key contributions are: (1) a robust study and bibliographic review of BI and the ETL process; (2) the design and development of a general-use tool called \dod{}, using state-of-the-art messaging, cluster computing tools and in-memory databases; (3) application of \dod{} to the steel sector.

The remainder of this work is organized as follows. First, we discuss related work in \strc{sec:rw}. Then, \strc{sec:design} presents \dod{} and the assumptions under which it has been designed, detailing its implementation and optimization. We evaluate performance, scalability, and fault-tolerance in \strc{sec:eval}. And finally, we summarize our main findings and propose future work in \strc{sec:con}.
\section{Related Work}
\label{sec:rw}



Due to the increasing pressure on businesses to perform decision-making on increasingly short time frames, data warehouses are required to be updated in real-time or on the shortest possible interval, not waiting for lower workload periods. This requirement leads to the concept of \nrt{} ETL, a challenging and important topic of research, whose primary definitions, problems and concepts were defined by Vassiliadis and Simitsis \cite{vassiliadis2009near}.

Other emerging concepts related to \nrt{} ETL are active data warehouses \cite{thalhammer2001active}, \cite{karakasidis2005etl} and real-time BI \cite{azvine2006real}, \cite{sahay2008real}, \cite{nguyen2005sense}. Both describe a new paradigm for business intelligence, in which data is updated in \nrt{} and both decision-making process as well as actions are performed automatically.

Both  Wibowo \cite{wibowo2015problems} and Sabtu et al. \cite{sabtu2017challenges} identify problems and challenges for developing \nrt{} ETL systems. Along with Ellis \cite{ellis2014real}, they provide some directions within this topic for researchers and developers, proposing possible solutions to each challenge. According to Wibowo \cite{wibowo2015problems}, there are problems on each stage of the ETL process that need to be handled to achieve \nrt{} ETL. Namely, data source overload, integration of heterogeneous sources, handling master data, and need for computational power to perform prompt operations.


Ellis \cite{ellis2014real} explains that dealing with \nrt{} ETL requires three key features to be addressed at each ETL phase: high availability, low latency, and horizontal scalability, which ensure that data will flow and be available constantly, providing up-to-date information to the business. They also enable ETL to always improve performance when needed by adding more servers to the cluster. Sabtu et al. \cite{sabtu2017challenges} also point out the challenges that \nrt{} solutions faces but focus on multiple and heterogeneous data sources, data backups, inconsistency, performance degradation, data source overload and master data overhead.

\strt{nrt-etl-ps} sums up these solutions and related problems, also described in detail below.

\begin{table}[H]
\centering
\caption{Near real-time ETL problems and solutions.}
\label{nrt-etl-ps}
\begin{tabular}{ll}
\multicolumn{1}{c}{\textbf{Problem}} & \multicolumn{1}{c}{\textbf{Solution}} \\ \hline
Multiple data source integration     & Stream Processing                     \\
Heterogeneous data source             & Stream Processing                     \\
Backup data                          & Log-based CDC                         \\
Performance degradation              & Log-based CDC                         \\
Data source overload                 & Update significance; Record change    \\
Master data overhead                 & Master data repository; join optimization               
\end{tabular}
\end{table}

Mesiti et al. \cite{mesiti2016streamloader} take advantage of the Stream Processing paradigm, where processors can work concurrently, to integrate multiple and heterogeneous sensor data sources. This work takes the integration challenge to a more severe domain, where heterogeneity and multiplicity of data sources are accentuated: sensors as data sources. Naeem et al. \cite{naeem2008event} address \nrt{} ETL by proposing an event-driven \nrt{} ETL pipeline based on push technology and with a database queue. They also tried to minimize master data overhead by designing a new approach that manages them during the transformation stage: they divided data into master data, which are more static, and transactional data, which changes more often, and stored that master data on a repository. This strategy made its use more efficient during the transformation step.

Zhang et al. \cite{zhang2015task} proposes a new framework to process streaming data in healthcare scientific applications. Their framework (1) enables integration between data from multiple data sources and with different arrival rates and (2) estimates workload so it can plan for computational resources to scale appropriately. They extended the Hadoop \cite{dean2008mapreduce} Map-Reduce framework to address the streaming data varied arrival rate. To estimate the unknown workload characteristics, they propose two methods to predict streaming data workload: smoothing and Kalman filter.

Waas et al. \cite{waas2013demand} propose a different approach for ETL with streaming. They first import raw records from data sources and only proceed with transformation and data cleaning when requested by reports, turning ETL into ELT. Therefore, data is transformed on-demand processing different data sets at different moments as required. To achieve this capability, they developed a monitoring tool to alert all connected components about new incoming data.  

Many works focused on the master data overhead problem, where joins between the data stream and master data can lead to performance bottlenecks. To minimize these bottlenecks, they proposed optimizations strategies to these joins \cite{polyzotis2007supporting, polyzotis2008meshing, bornea2011semi, naeem2010r}.

As for data source overhead, Jain et al. \cite{jain2012refreshing} compared CDC proposals using two metrics to identify data source overload.  They concluded that log-based CDC was the most efficient, having minimal impact on database performance and minimizing data loss.

The above-mentioned publications propose strategies to overcome the challenges related to the three main features of \nrt{} ETL solutions (high availability, low latency and horizontal scalability). Each one focusing on a specific problem and not all of them at once. However, the Stream Processing paradigm appears as a common denominator among most solutions, which is consistent with \nrt{} ETL requirements.

In this stream-based application context, multiple stream processing frameworks facilitate the development of such applications or are used directly to solve \nrt{} ETL requirements. Most of these frameworks are based on the record-at-a-time processing model, in which each record is processed independently. Examples of stream processors frameworks that use this model are Yahoo!’s S4 \cite{neumeyer2010s4} and Twitter’s Storm \cite{toshniwal2014storm}.

In contrast to this record-at-a-time processing model, another possibility is to model it as a series of small deterministic batch computations, as proposed by Zaharia et al. \cite{zaharia2012discretized} as the Spark Streaming framework. This way, among others benefits, integration with a batch system is made easy and performance overhead, due to record-at-a-time operations, is avoided. Flink \cite{carbone2015apache} is currently competing with Spark Streaming as the open source framework for heavyweight data processing. It also merges, in one pipeline, stream and batch processing and it has features such as flexible windowing mechanism and exactly once semantics.

Besides these open source frameworks, there are those offered by cloud providers as services such as Google Dataflow \cite{googledataflow} and Azure Stream Analytics \cite{azurestreamanalytics}. By using these services, it is possible to avoid the installation and configuration overhead and the resources allocation to horizontal scalability gets simpler. 

These open source frameworks and stream processing services are both designed for general use. Due to this one size fits all architecture, they lack strategies and optimizations that are used by \dod{}. In addition, while the above-mentioned publications propose solutions to a specific problem or challenge, \dod{} tries to combine them in a single tool.

\section{\dod{}}\label{sec:design}

\dod{} relies on an on-demand data stream pipeline with a distributed, parallel and technology-independent architecture. It uses Stream Processing along with an in-memory master data cache to increase performance, a buffer to guarantee consistency of join operations on data with different arrival rates, and a unified programming model to allow it to be used on top of a number of Stream Processing frameworks. Besides, our approach takes advantage of (1) data partitioning to optimize parallelism, (2) data filtering, to optimize data storage on \dod{}'s \imc{} and (3) log-based CDC, to process data on-demand and with minimum impact on the source database. Therefore, \dod{} has: minimum impact on the source database during extraction, works in \nrt{}, can scale to respond to data and throughput growth and can work with multiple Stream Processing frameworks.

The insights that enable \dod{} to achieve the features mentioned above are: {\em On-demand data stream}---as data changes on the source database, they are processed and loaded into the target database, creating a stream of data where the ETL process handles, in a feasible time frame, only the minimum amount of necessary data. {\em Distributed \& parallel}---perform all steps in a distributed and parallel manner, shortening processing time and enabling a proper use of the computer resources. {\em \imc{}}---perform data transformations with no look-backs on the source database, providing all required data to execute calculations in the \imc{} which avoids expensive communications with the data source. {\em Unsynchronized consistency}---a buffer guarantees that data with different arrival rates can be joined during transformation. {\em Unified programming model}---a programming model used to build data pipelines for both batch and streaming in multiple Stream Processing frameworks.

\dod{} merges into a single solution multiple strategies and techniques to achieve near real-time ETL that have are have never been integrated to this extent: log-based Change Data Capture (CDC), stream processing, cluster computing, and an in-memory data store along with efficient data partitioning (c.f. \strc{sec:rw}). We note that although these techniques have indeed been used before (e.g., in \cite{mesiti2016streamloader,naeem2008event,zhang2015task,jain2012refreshing})) they have not been previously integrated into a single solution. By synergistically combining these strategies and techniques \dod{} can achieve \nrt{} ETL.

\subsection{Architecture}

\dod{} has the following workflow (\strf{arch}): (1) it tracks changes on each source system's database table and (2) sends these changes as messages to a (3) message queue, following a preconfigured partitioning criteria. Then, (4) these messages are pulled from the message queue and sent to the \imc{} or (5) transformed to the target reporting technology format, and, finally, (6) loaded into the target database. \dod{}'s workflow can be grouped into two central modules: \ct{} and \spr{}. 

\begin{figure}[H]
\centering
\includegraphics[scale=0.6]{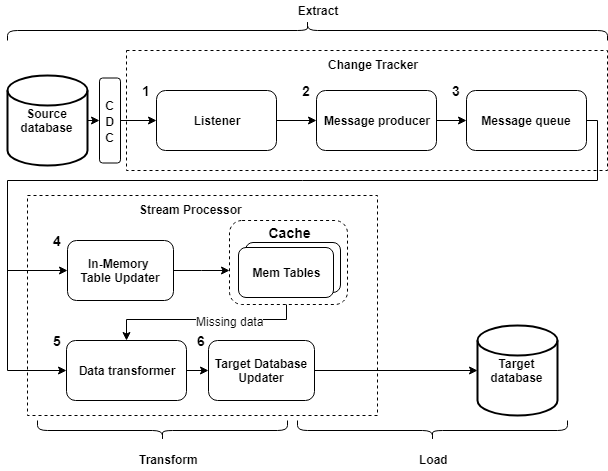}
\caption{\dod{} workflow step by step.}
\label{arch}
\end{figure}

All steps depend on configuration parameters to work properly. Thus, during \dod{}'s deployment, it is imperative to go through a configuration process, where decisions are made to set the following parameters: {\em tables to extract}---define which tables will have data extracted from; {\em table nature}---from the defined tables, detail which ones are operational (constantly updated) and which ones are master data (more static); {\em table row key}---column, from each table, that contains each row's unique identifier; {\em business key}---column, from each table, that contains the values used to partition or filter data by a domain-specific key.

\subsubsection{\ct{}}

The \ct{} makes available to the \spr{} module, at the same time that events occur, any data altered and added to the source database. This module makes use of CDC, a standard database pattern that contains all operations carried out over time and their respective values. Thus, whenever a record is inserted, altered, or removed from a table, the CDC writes that event, together with all of the record's information.

CDC can take many forms, ranging from log files to structured tables in the database. They can be generated either by the database itself or by an external system. In this work, to be adherent to  Jain et al.'s recommendations \cite{jain2012refreshing}, log-based CDC was used. However, as explained later, the CDC reading step in \dod{} is performed by the \lstr{}; thanks to \dod{}'s modular architecture, it is possible to create different \lstr{}s to support different CDC implementations. 

The \ct{} works in three stages called \lstr{}, \mspr{}, and \msq{}. The \lstr{} stage listens to the CDC changes and, with each new record, extracts its data to be further processed by \mspr{}. This stage was built to extract data independently among tables, allowing it to be parallelized. As a prerequisite, then, \lstr{} expects the CDC log to support this feature. The \lstr{} step has minimum impact on the source database's performance due to two factors: (1) only new and altered records are extracted at each execution, representing a lower data volume, and (2) queries are performed in log files only, which takes the pressure off the database and production tables. As this step has been designed to be decoupled from the others, it can be rebuilt to meet the specific characteristics of the source database, such as different technologies and CDC implementations.

The \msq{} works as a message broker and uses the publish/subscribe pattern with partitioning features, in which each topic is partitioned by its messages key.
On \dod{}, a topic contains all insertions, updates and deletions of a table. The topic partitioning criteria vary by the table nature that the topic represents (master or operational). When dealing with master data tables, each topic is partitioned by its respective table unique row identifier and, when dealing with operational data, each topic is partitioned by the business key. This partitioning occurs at the \mspr{}, before it publishes each extracted data, based on the aforementioned configuration parameters. Therefore, \mspr{} builds messages from data extracted by the \lstr{} and publishes them in topics on the \msq{} according to the preconfigured parameters. These two partitioning strategies (by row key and by business key) have the following objectives:

{\em Row key}: since the table unique row identifier is the topic partition key, \msq{} guarantees that the consumption of messages from a specific table row will happen in the right sequence. By calling the last message from each partitioning key (row id), it is possible to reconstruct the most recent table snapshot. In other words and, as shown in \strf{tts}, each topic contains all insertions, updates and deletions from a table and it is partitioned by its id column. The last value from all partitions key represents the current table state.

{\em Business key}: the \spr{} transformation process is parallelized based on the number of partitions defined by operational topics and their business keys. Therefore, the partitioning criteria has a direct impact on \dod{}'s performance. In this sense, it is paramount to understand in advance the extracted data and the nature of operations that will be performed by \spr{}. The goal is to figure out the partitioning criterion and its key, because they may vary depending on the business nature.


\subsubsection{\spr{}}

The \spr{} receives data from the \lstr{} by subscribing to all topics published on \msq{}, receiving all recently changed data as message streams. \spr{} comprises three steps: (1) \imtu{}, (2) \dtt{} and (3) \tdu{}. 

{\em \imtu{}} prevents costly look-backs on the source database by creating and continuously updating distributed in-memory tables that contains supplementary data required to perform a data transformation. Data flowing from topics representing master data tables goes to \imtu{} step. \imtu{} only saves data related to the business keys assigned to its corresponding \spr{} node, filtering messages by this key. By doing so, only data from keys that will be processed by each node are saved in its in-memory table, taking off pressure from memory resources. In case of node failures, data loss or both, \spr{} retrieves a snapshot from \msq{} and repopulates in-memory tables that went down. This is possible due to the way each \msq{} master data topic is modeled: it is partitioned by the table's unique row identifier, allowing \imtu{} to retrieve, from \msq{}, an exact snapshot of this topic table.


{\em \dtt{}} receives data and performs operations to transform it into the required BI report format. \dtt{} can run point-in-time queries on the in-memory tables to fetch missing data necessary to carry out its operations, joining streaming and static data efficiently. Each partition is processed in parallel, improving performance. The operations executed in \dtt{} rely on the operators of the cluster computing framework (e.g., map, reduce, filter, group) and are business-dependent. Like \imtu{}, not all messages go through this module, only messages from tables configured as operational. In the event of master data arriving after operational data (master and operational messages are sent to different topics and by different \lstr{} instances), \dtt{} uses a buffer to store this late operational message for late reprocessing. At each new operational message, \dtt{} checks the buffer for late messages and reprocesses them, along with the current one. To optimize performance, \dtt{} only reprocesses buffer messages with transaction dates older than the latest transaction date from the \imc{}, which avoids reprocessing operational messages that still have no master data As shown in more details in \strc{sec:eval}, \dod{} performance is highly dependant on the data model and transformation operations complexity. That is, the data and how it is transformed, in this case by \spr{}, will determine \dod{} processing rate.

{\em \tdu{}} translates the \dtt{}'s results into query statements and then loads the statements into the target database. For performance, the loading process also takes place in parallel and each partition executes its query statements independently.

\subsection{Scalability, Fault-tolerance and Consistency}

\dod{} takes advantage of the adopted cluster computing framework's and the message queue's native support for scalability and fault-tolerance. For the former, both \dod{} modules can handle node failures, but with different strategies: while \ct{} focuses on no message loss, \spr{} concentrates on no processing task loss. As for scalability, both depend on efficient data partitioning to scale properly. 

Despite these inherited capabilities, some features had to be implemented on \spr{} so fault-tolerance and scalability could be adherent to \dod{}'s architecture: the \imc{} and the \omb{} have to be able to retrieve data from new nodes or from failed nodes. Regarding the \imc{}, we implemented a trigger that alerts \imtu{} when \dtt{}'s assigned business keys changes. On a change event, \imtu{} resets the \imc{}, dumps the latest snapshot from \msq{} and filters it by all assigned business keys. By doing so, the \imc{} keeps up with \spr{} reassignment process on failure events or even when the cluster scales up or down. As for the \omb{}, it uses a distributed configuration service, that already comes with \msq{}, to save its values so, in case of a node failure, other \spr{} instances can keep up with reprocessing. That is, \omb{} saves all operational messages with late master data and, at each new message, \dtt{} tries to reprocess this buffer by checking at the \imc{} if its respective master data arrived, solving the out-of-sync message arrival problem.

Since \dod{} focuses on delivering data to information systems, not operational mission-critical systems, transactional atomicity was not a requirement and it was left out of the scope of this work: rows of different tables added/altered in the same transaction can arrive at slightly different time frames. However, due to its late operational messages buffer, \dod{} can guarantee that only data with referential integrity will be processed and that those that are momentarily inconsistent will be eventually processed. As stated before, \dtt{} reprocesses all messages from the \omb{} when all needed data arrives on the \imc{}.

\subsection{Implementation}

\dod{} is a tool with many parts: CDC, \lstr{}, \msq{}, \spr{}. While the \lstr{} was built from scratch, CDC and \msq{} are simply out of the shelf products, and \spr{} is a Stream Processing Framework with customizations and optimizations built on top of a unified programming model. Its implementation and used technologies are explained next.

All data exchanged between \dod{} modules are serialized and deserialized by the Avro system \cite{apacheavro}. The \msq{} role was performed by Kafka \cite{kreps2011kafka}, an asynchronous real-time message management system, whose architecture is distributed and fault-tolerant. Due to Kafka's dependency on Zookeeper \cite{hunt2010zookeeper}, it was also used on \dod{}. 

\spr{} was implemented with Beam \cite{apachebeam}, a unified programming model, which allows it to be used on top of Stream Processing frameworks such as Spark Streaming and Flink. Its steps, \dtt{}, \imtu{} and \tdu{}, were all encapsulated together to make communication between them easy. \dtt{} takes advantage of the already deployed Zookeeper to store its late operational messages buffer. It does so to guarantee that, in any failure event, another \spr{} node could keep processing those late messages.

Regarding the \imc{}, H2 \cite{h2db} was used and deployed as an embedded database on the \spr{} application. To support \dod{}'s needs, H2 was configured to work in-memory and embedded so, for each Spark worker, we could have an H2 instance with fast communication. Our prototype and its modules are publicly available.\footnote{https://github.com/gustavo-vm/dod-etl}

Since \dod{} modules were designed to be decoupled, each one can be altered and replaced without impacting the others. Adding to this its technological-independent features, all selected technologies on each module can be replaced, provided that its requirements are satisfied.

\section{Evaluation}\label{sec:eval}




We used, as a case study, a large steelworks and its BI processes. In this case, a relational database, powered by both its production management system and shop floor level devices, was used as the data source. This steelworks has the following characteristics: total area of 4,529,027\,m$^2$, a constructed area of 221,686\,m$^2$, 2,238 employees, and it is specialized in manufacturing Special Bar Quality (SBQ) steel. 



This steelworks uses OLAP reports \cite{codd1993providing} as its BI tool. \dod{} was used to provide \nrt{} updates to these reports, which were unavailable; prior to \dod{}'s deployment reports were updated twice a day. \dod{}'s purpose was to extract data from the source database and transform them into OLAP's expected model, known as \emph{star schema} \cite{giovinazzo2000object}. This transformation involves calculations of Key Process Indicators (KPIs) of this steelworks' process. For this case study, the Overall Equipment Effectiveness (OEE) \cite{Stamatis201006}, developed to support Total Productive Maintenance initiatives (TPM) \cite{ljungberg1998measurement}, and its related indicators (Availability, Performance, and Quality) were chosen as the steelworks KPIs.

TPM is a strategy used for equipment maintenance that seeks optimum production by pursuing the following objectives: no equipment breakdowns; no equipment running slower than planned; no production loss.

OEE relates to TPM initiatives by providing an accurate metric to track progress towards optimum production. That is, the following KPIs can quantify the above three objectives: {\em availability}---measures productivity losses, tracking the equipment downtime vs. its planned productive time; {\em performance}---tracks the equipment actual production speed vs. its maximum theoretical speed; {\em quality}---measures losses from manufacturing defects. These three indicators together result in the OEE score: a number that provides a comprehensive dimension of manufacturing effectiveness. 





\dod{} extracted only the data needed to calculate these indicators. For this industry context, we grouped them into the following categories: {\em production data}---information of production parts; {\em equipment data}---equipment status information; {\em quality data}---produced parts quality information. During the \dod{} configuration process, two decisions were made: we defined the nature of each table (operational and/or master data) and decided which table column would be considered as the \spr{} business partitioning key. Regarding the table nature, we considered the production group as operational and equipment and quality as master data. Due to this decision, all equipment and quality data sent to \spr{} will be stored on its \imc{} while production data will go straight to the \dtt{} step of the \spr{}.

As for the business key, we considered the production equipment unit identifier, since all KPIs are calculated for it. We set, then, the column that represents and stores the related equipment unit code on each table as the business key in the configuration. This column will be used by operational topics for partitioning and by the \imc{} as filter criteria.




For this industry context, \dtt{} splits data as a requisite to support OLAP multidimensional reports. For the above mentioned KPIs, we defined the splitting criteria as its intersections in the time domain, in which the lowest granularity represents the intersection between all the data obtained for the equipment unit in question: Production, Equipment status and Quality. \strf{part} shows an example of this intersection analysis and data splitting. In this case, \dtt{} searches for intersections between equipment status and production data and breaks them down, generating smaller data groups called fact grain. In the \strf{part} example, these fact grains can be divided in two groups: (1) equipment with status "off" and (2) equipment with status "on" with production. As stated before, after the splitting process is completed, \dtt{} performs the calculations.

\begin{figure}[H]
\centering
\includegraphics[scale=0.55]{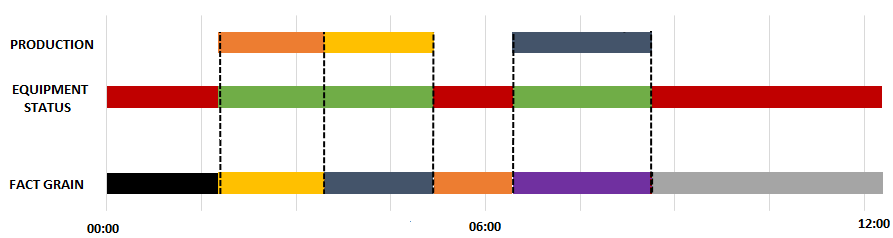}
\caption{Data splitting working on metals industry context.}
\label{part}
\end{figure}


\subsection{Experiments}\label{sec:exp}



To evaluate our \dod{} prototype's performance, we used the Spark Streaming framework \cite{zaharia2012discretized} as the baseline. We generated a synthetic workload, simulating the data sent by the steelworks equipment, and executed Spark with and without \dod{} on top of it. We have also performed experiments to check if \dod{} achieved Ellis \cite{ellis2014real} key features (high availability, low latency and horizontal scalability) and, as a differential of our work, we executed \dod{} with production workloads from the steelworks to check its behavior in a complex and realistic scenario.

In sum, we evaluated (1) \dod{} vs Baseline, checking how Spark performs with and without \dod{} on top of it; (2) horizontal scalability, analyzing processing time when computational resources are increased; (3) fault tolerance, evaluating \dod{}'s behavior in the event of failure of a compute node; (4) \dod{} in production, comparing its performance against real workloads and complex database models from the steelworks.

We used Google Cloud and resources were scaled up as needed, except for the fifth experiment that used the steelworks's computing infrastructure. The following hardware and configurations were used: {\em database} -- MySQL (with CDC log activated) deployed on an 8-core 10\,GB memory instance; {\em sampler} -- 20-core 18\,GB instance; {\em \ct{}} -- 10-core 18\,GB instance; {\em \msq{}} -- seven instances, three for Kafka brokers and three for Zookeeper, with one core and 2\,GB of memory each; {\em \spr{}} -- we deployed Spark in standalone mode with one master node and multiple worker nodes, which varied from one to twenty instances depending on the experiment (all \spr{} instances had one core and 2\,GB of memory each). All Google Cloud Platform instances had Intel Haswell as their CPU platform and hard disk drives for persistent storage. To represent the three data categories cited before, we used one table per data model group on experiments 1, 2 and 3. Regarding the fourth experiment, we used a more complex data model based on the ISA-95 standard~\cite{InternationalSocietyofAutomation2001}.

Also for experiments 1, 2 and 3, as mentioned above, we built a sampler to insert records on each database table. This sampler generates synthetic data, inserting 20,000 records at each table, simulating  the steelworks operation. To minimize the impact of external variables in these experiments, the \lstr{} started its data extraction after the sampler finished its execution. To avoid impact on the results, \ct{} extracted all data before the execution of \spr{}, so \msq{} could send data at the same pace as requested by \spr{}. Since \lstr{} depends on the CDC log implementation, its data extraction performance also depends on it. We used MySQL as the database and its binary log as the CDC log.

\subsubsection{\dod{} vs Baseline}

Since \dod{} is comprised of out-of-the-shelf components, each module following a decoupled architecture. To evaluate \dod{}'s performance, each module needed to be analyzed separately.

As said before, \lstr{} is highly dependant on the used database and CDC implementation and its performance is tied to them. Therefore, \lstr{} has the CDC throughput as its baseline. Since the complete flow from writing to the database to copying it to the binary incurs a lot of I/O, \lstr{} will always be able to read more data than CDC can provide.

Since \msq{} is an out-of-the-shelf product and can be replaced by any other product, provided that its requirements are satisfied, it can keep up with new benchmarks. As for now, \mspr{} and \msq{} are instances of Kafka producers and Kafka brokers, respectively. Kreps et al. \cite{kreps2011kafka} already demonstrated its performance against other messaging systems.

The \spr{} includes substantial customizations (\dtt{}, \imtu{} and \tdu{}) on top of the Spark Streaming framework. We evaluated its performance executing the same workload against Spark Streaming with and without \dod{}. We used a fixed number of Spark worker nodes (\strt{tab-exp-re}): a cluster with ten nodes.


While \dod{} was able to process 10,090 records per second, Spark Streaming alone processed ten times fewer records (1,230 records/s). In order to understand in detail this performance difference, we analyzed the Spark job execution log, that shows the processing time in milliseconds of each Spark Worker task (\strf{fig-mem-perf}). Note that \dod{} has an initialization overhead at each Spark worker when it is started and when a new key or partition is assigned. Regarding the used dataset and infrastructure, this initialization overhead costs 40 seconds for each Worker.

Throughout these experiments, we were able to demonstrate that \dod{} customizations make Spark significantly faster. Although it has an initialization overhead, due to the \imtu{} data dump from \msq{}, it is minimal and negligible considering the volume of processed messages.

\begin{figure}[H]
\centering
\includegraphics[scale=0.65]{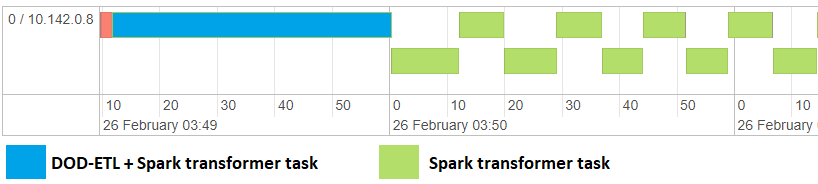}
\caption{In-memory cache initialization overhead.}
\label{fig-mem-perf}
\end{figure}

\subsubsection{Scalability}


We evaluated the scalability of \dod{}'s \ct{} and \spr{} modules. \lstr{} scalability, as stated before, also depends on the CDC log implementation. We performed two different experiments to evaluate \lstr{}'s performance: (1) {\em Inserted data on extracted tables only}, where insertions were made only in databases that we were extracting data, so the number of inserted and extracted tables increased at the same time (from 1 to 16); and {\em Fixed number of inserted tables}, where insertions were made in a fixed number of tables (16) and the number of extracted tables was increased (from 1 to 16).

As shown in \strf{fig-lstr-scalab}, where number of records inserted per second was plotted against the number of tables, \lstr{} had different results on each experiment: When inserting data on extracted tables only, \lstr{}'s performance increased as a sublinear function and then saturated at 18,600 records per second for eight tables. When using a fixed number of tables, it increased linearly until it also saturated when extracting simultaneously from eight tables, with a throughput of 10,200 records per second. This behavior is directly related to MySQL's CDC implementation, where it writes changes from all tables on the same log file, so each \lstr{} instance had to read the whole file to extract data from its respective table. Throughput is higher in the first experiment compared to the second experiment because of the difference in log file size: while in the fixed insertions experiment the CDC log file had a fixed size of 320,000 records, the varying insertion experiment log file varied from 20,000 records to 320,000 records. Therefore, going through the whole log file took less time until it matched at 16 tables. Regarding the saturation point, this happened due to MySQL performance and we conjecture it will vary across different databases and CDC implementations.

\begin{figure}[H]
\centering
\includegraphics[scale=0.8]{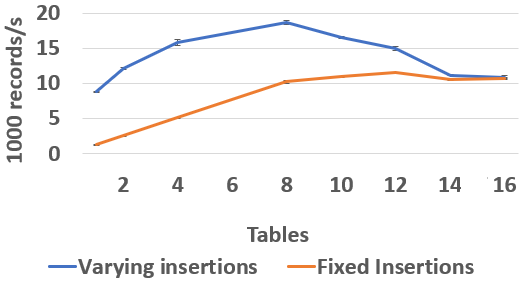}
\caption{Scalability: \lstr{} experiment result.}
\label{fig-lstr-scalab}
\end{figure}


As already stated, \mspr{} and \msq{} are instances of Kafka producers and Kafka brokers, respectively. Kreps et al. \cite{kreps2011kafka} already demonstrated that each producer is able to produce orders of magnitude more records per second than \lstr{} can process. Regarding its brokers, their throughput is dictated more by hardware and network restrictions than by the software itself, also enabling it to process more records.




To evaluate \spr{}'s scalability, we varied the number of available Spark worker nodes from one to twenty and fixed the number of partitions on the operational topic at twenty. To maximize parallelization, the number of equipment units (partition keys) from the sampled data followed the number of topic partitions: sampled data contained 20 equipment unit identifiers, used as partition keys. We used the Spark Streaming Query Progress' metric \emph{average processed rows per second} at each of its mini batch tasks. As shown in \strf{fig-spr-scalab}, where the number of processing records per seconds was plotted against the number of Spark Workers.


\begin{figure}[H]
\centering
\includegraphics[scale=0.7]{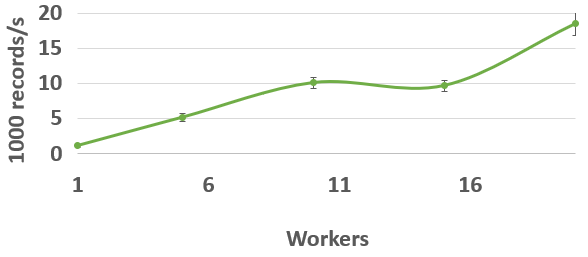}
\caption{Scalability: \spr{} experiment result.}
\label{fig-spr-scalab}
\end{figure}



\dod{} modules are scalable: both \ct{} and \spr{} can absorb growth by adding more computational resources, despite their difference in throughput and scalability factor. While \ct{} scales proportionally to the number of tables, \spr{} scales with the number of partition keys on operational tables.

Regarding \ct{}, we saw that \lstr{} and \mspr{} can process tables independently and that it can scale up as the process incorporates new ones, provided that the database CDC log supports it. As for \msq{}, it also scales linearly but based on multiples variables: the number of available brokers, extracted tables (topics), partitions, and keys.

\spr{}'s scalability is proportional to the number of partitions at the operational table topics and the number of partitioning keys that, on this steelworks case, are the total number of production equipment units. Data partitioning plays a vital role here, so, it is imperative to understand functionally and in advance all extracted data in order to find partitioning criterion and its key, which varies from business to business. Since \msq{} supports throughput orders of magnitude higher than \lstr{} and \spr{}, it is possible to integrate multiple database sources and use multiple \spr{} instances, each performing different transformations.



\subsubsection{Fault Tolerance}

We have executed \dod{} in a cluster with five worker nodes and, midway through the experiment, we shut down two of the worker nodes to evaluate \dod{}'s fault tolerance. We measured the rate of processed messages before and after the shutdown and performed a data consistency check on the result. We used the same performance metric as the scalability experiment (\strt{tab-exp-re}).


\spr{} went from processing 5,060 messages per second to 2,210, representing a processing rate decrease of 57\%. After each execution, we checked the consistency of all messages processed by this module and did not find any error: it processed all messages correctly, albeit at a reduced rate. This result indicates that \dod{} is fault tolerant, which significantly increases its robustness.

While the number of available clusters was changed from 5 to 3, a 40\% decrease, the performance decrease was more significant (57\%). By analyzing the Spark execution log, we found that the \imc{} also impacts fail-over performance: when a node becomes unavailable and a new one is assigned, the \imc{} has to dump all data from the newly assigned partition keys, which impacts performance.

Since the \ct{} and the \spr{} were built on top of Kafka and Spark, respectively, both modules can resist node failures. Due to their different purposes, each module uses distinct strategies: while Kafka can be configured to minimize message loss, Spark can be configured to minimize interruption of processing tasks.

\subsubsection{\dod{} in production}


We executed \dod{} with real workloads from the steelworks. This workload is generated by both a production management system and shop floor level devices and its data model is based on the ISA-95 standard, where multiple tables are used to represent each category (production, equipment and quality). We compared \dod{} results on previous experiments (\strt{tab-exp-re}), where synthetic data was used with a simpler data model (a single table for each category of data), with \dod{} executing real and complex data.

\begin{table}[H]
\large
\centering
\caption{Experiments results (records/s).}
\label{tab-exp-re}
\begin{tabular}{cccccc}
\multicolumn{2}{c}{\textbf{Baseline}}         & \multicolumn{2}{c}{\textbf{Fault Tolerance}}        & \multicolumn{2}{c}{\textbf{Production}}       \\ \hline
\dod{}                 & Spark                 & Normal                   & Failure                  & Simple Wkld                    & Real Wkld                 \\ \hline
\multicolumn{1}{r}{10,090} & \multicolumn{1}{r}{1,230} & \multicolumn{1}{r}{5,063} & \multicolumn{1}{r}{2,216} & \multicolumn{1}{r}{10,090} & \multicolumn{1}{r}{230}
\end{tabular}
\end{table}

Both synthetic and production experiments used the same configuration: a cluster with ten Spark worker nodes. While \dod{} can process 10,090 records per second for the data source with simple model complexity, this number decreases to 230 records per second for the complex data model. It is possible to state, then, that data model complexity impacts directly on \dod{} performance. Since it depends on the \imc{} to retrieve missing data, when this retrieval involves complex queries, this complexity impacts on the query execution time and, therefore, on \dod{} performance.

This steelworks currently uses an ETL solution to perform the same operation performed by \dod{}. It adopts a sequential batch approach, comprises a series of procedures ran within the relational database server, and relies on a twelve core /32 GB memory server. While \dod{} takes 0.4 seconds to process 100 records, this solution takes one hour. Although it is not a a fair comparison (a streaming distributed and parallel tool vs a batch and legacy solution), it is important to demonstrate that \dod{} can be successfully used in critical conditions and to absorb the steelworks data throughput, providing information in \nrt{} to its BI tools.

Considering the author's experience in developing mission critical manufacturing systems and its knowledge in the ISA-95 standard, his opinion regarding these systems data modeling is that the drawbacks of using a standardized and general data model, that seeks a single vision for all types of manufacturing processes, far outweigh the benefits. Manufacturing systems that use generalized data models get way more complex when compared with process-specific models. These systems performance, maintenance and architecture get severely impacted in order to provide a generic model. 

Therefore, in this industry context, a more straightforward data model could be used in production management systems and shop-floor without drawbacks. With this, \dod{} would perform even better, in addition to the factory systems.

\section{Conclusion}\label{sec:con}


\dod{}'s novelty relies on synergistically combining multiple strategies and optimizations (that were previously only used separately) with an on-demand data stream pipeline as well as with a distributed, parallel, and technology-independent architecture.



ETL systems need to have three key features to work in \nrt{}: high availability, low latency, and scalability. \dod{} has been able to achieve all three key features and address these challenges by combining log-based Change Data Capture (CDC), stream processing, cluster computing, an in-memory data store, a buffer to guarantee join consistency along with efficient data partitioning, and a unified programming model.

We have been able to demonstrate, by performing independent experiments on each of its main modules, that \dod{} strategies and optimizations reduce ETL run time by a factor of ten, outperforming unmodified Stream Processing frameworks. Through these experiments, we showed that \dod{} achieves these results without sacrificing scalability and fault-tolerance. We have also found that data source model complexity heavily impacts the transformation stage and that \dod{} can be successfully used even for complex models.

Due to its technology-independence, \dod{} can use a number of Stream Processor frameworks and messaging systems, provided that its requisites are satisfied. This allows \dod{} to adapt and evolve as new technologies arrive and avoid technology lock-ins.

Instantiating \dod{} requires customizing the \dtt{} step: each required transformation is translated as Spark operators which, in turn, are compiled as a Java application. This requires \dod{}'s users to know how to program, restricting its use. To overcome this, on future work, \dod{} will be adapted to integrate a friendly user interface with an intuitive and visual configuration of \dtt{} transformation operations.

Also on future work, we will study the impact on \dod{} performance when lightweight Stream Processing frameworks are used, such as Kafka Streams and Samza, by performing new experiments. By doing so, we will be able to compare and evaluate the trade-offs between these two types of frameworks (lightweight vs heavyweight) and its impact on \dod{} proposed strategies and optimizations.

In sum, through this work, we were able to achieve \nrt{} ETL by combining multiple strategies and technologies, to propose a general use tool and to evaluate it in the metals industry context.


\bibliographystyle{abbrvnat}
\bibliography{bib}

\end{document}